\documentclass[showpacs,preprintnumbers,amsmath,amssymb]{revtex4}

\usepackage{graphicx}
\usepackage{dcolumn}
\usepackage{bm}

\newcommand{\be}{\begin{equation}}
\newcommand{\ee}{\end{equation}}
\newcommand{\bea}{\begin{eqnarray}}
\newcommand{\eea}{\end{eqnarray}}

\newcommand{\integer}{\relax{\rm I\kern-.18em N}}

\begin{document}

\title{Density resummation of perturbation series in a pion gas \\to leading order in chiral perturbation theory}


\author{M. I. Krivoruchenko$^{\;1,2)}$, C. Fuchs$^{\;2)}$, B. V. Martemyanov$^{\;1)}$, }
\author{Amand Faessler$^{\;2)}$}

\affiliation{
$^{1)}$Institute for Theoretical and Experimental Physics$\mathrm{,}$
B. Cheremushkinskaya 25\\ 117259 Moscow, Russia \\
$^{2)}$Institut f\"{u}r Theoretische Physik$\mathrm{,}$ Universit\"{a}t T\"{u}bingen$\mathrm{,}$
Auf der Morgenstelle 14\\ D-72076 T\"{u}bingen$\mathrm{,}$ Germany}

\begin{abstract}
The mean field (MF) approximation for the pion matter, being equivalent to the 
Leading ChPT order, involves no dynamical loops and, if self-consistent, produces 
finite renormalizations only.
The weight factor of the Haar measure of the pion fields, entering the path
integral, generates an effective Lagrangian $\delta \mathcal{L}_{H}$ which is
generally singular in the continuum limit. There exists one parameterization
of the pion fields only, for which the weight factor is equal to unity and $%
\delta \mathcal{L}_{H}=0$, respectively. This unique parameterization ensures 
selfconsistency of the MF approximation. We use it to calculate thermal Green 
functions of the pion gas in the MF approximation as a power series over the 
temperature. The Borel transforms of thermal averages of a function 
$\mathcal{J}(\chi ^{\alpha }\chi ^{\alpha })$ of the pion fields $\chi ^{\alpha }$ 
with respect to the scalar pion density are found to be $\frac{2}{\sqrt{\pi }}\mathcal{J}(4t)$.
The perturbation series over the scalar pion density for basic characteristics of the 
pion matter such as the pion propagator, the pion optical potential, the scalar quark 
condensate $<{\bar{q}}q>$, the in-medium pion decay constant ${\tilde{F}}$, 
and the equation of state of pion matter appear to be asymptotic ones. These series 
are summed up using the contour-improved Borel resummation method. The quark scalar 
condensate decreases smoothly until $T_{max}\simeq 310$ MeV. The temperature $T_{max}$ 
is the maximum temperature admissible for thermalized non-linear sigma model at zero 
pion chemical potentials.
The estimate of $T_{max}$ is above the chemical freeze-out temperature $T\simeq 170$ MeV at RHIC
and above the phase transition to two-flavor quark matter $T_{c} \simeq 175$ MeV, 
predicted by lattice gauge theories. 
\end{abstract}


\maketitle


\section{Introduction}

\setcounter{equation}{0} 

Ultrarelativistic heavy-ion collisions allow to study phase diagram of QCD at high temperatures 
and small chemical potentials. Under such conditions, lattice gauge theories (LGTs) predict  
a crossover or a first-order phase transition into the deconfined phase, accompanied by 
the restoration of chiral symmetry. In two- and three-flavor LGTs, the critical temperature 
is estimated to be $T_{c} \simeq 175$ MeV and $T_{c} \simeq 155$ MeV with a 5\% systematic 
error \cite{KALAPE,AALIK,Aoki06}. These values are close to the chemical freeze-out temperature 
determined from statistical models by fitting particle yields in heavy-ion collisions \cite{BRMU01}. The chemical freeze-out temperatures $T=160\div 174$ MeV and $T=160\div 166$ MeV are extracted \cite{andronic06} at top SPS energies ($\sqrt{s_{NN}}=17.3$ GeV) and top RHIC energies ($\sqrt{s_{NN}}=200$ GeV), respectively. The boundary of the phase transition might apparently be already crossed.

In ultrarelativistic heavy-ion collisions pions are the most abundant 
particles \cite{BRMU01,BRAV02,andronic06}, so thermodynamic characteristics 
of a pion-dominated medium and in-medium pion properties are of current interest. 
At zero chemical potentials, light degrees if freedom, i.e. pions, give the dominant 
contribution to the pressure of hadron matter \cite{KMK}, while 
contributions of resonances like $\rho$- and $\omega$-mesons with masses $M \gg T_{c}$ 
are suppressed as $\frac{T}{M}\exp(-\frac{M}{T}) \ll 1$. According to the Gibbs' criterion,
the balance of pressure determines the critical temperature of the phase 
transition at fixed chemical potentials. The QCD phase transition can be driven therefore by pions i.e. by the lightest QCD degrees of freedom.

The energy density of a resonance gas is proportional to the small parameter 
$\exp(-\frac{M}{T}) \ll 1$ which can, however, be compensated by an exponentially large 
amount of resonances, as has been conjectured by Hagedorn \cite{hagedorn1,hagedorn2} 
and discussed recently e.g. in Ref. \cite{KART}. The contribution of resonances 
to the pressure is still suppressed in such models by a factor $\frac{T}{M} \ll 1$ where 
$M \sim 1$ GeV is a typical scale of resonance masses.

Pions are Goldstone particles and play a special role in the restoration of 
chiral symmetry. LGTs give an evidence that deconfinement and chiral phase transitions 
occur at the same critical temperature for zero chemical potentials \cite{KALA03}.

Chiral perturbation theory (ChPT) exploits the invariance of strong interactions under the $SU(2)_{L}\otimes SU(2)_{R}$ symmetry \cite{GL84,GL85}. It has been proven to be highly successful in phenomenological
descriptions of low-energy dynamics of pseudoscalar mesons \cite{MEIS,DOBA}. The pion matter created in relativistic heavy-ion collisions is charge symmetric \cite{BRMU01}, so the pion chemical potentials can be set equal to zero. The ChPT expansion in the vacuum runs over inverse powers of the pion decay constant $F=93$ MeV and, in the medium, additionally over the pion density (or temperature). Properties of the pion gas at finite temperatures, the associated chiral phase transition, and the in-medium modifications of pseudoscalar mesons within ChPT have extensively been studied \cite{GL87,GELE89,ELET93, ELET94,PITY96,BOKA96,JEON,TOUB97,dobado,KRIV,LOEWE}.

The optical potential of an in-medium particle yields the self-energy operator to the first order in the density. It is determined by the forward two-body scattering amplitude on particles of the medium. ChPT is suited for calculation of the pion self-energy beyond the lowest order in density, since ChPT has all multi-pion scattering amplitudes fixed. The forward scattering amplitudes of a probing pion scattered off $n$ thermalized pions describe $O(\rho ^{n})$ contributions to the in-medium pion self-energy operator. The leading ChPT amplitudes do not contain loops. This approximation is equivalent to the mean field (MF) approximation.

In this paper, we perform to the leading ChPT order the generalized Borel resummation of the asymptotic density series for the pion propagator, the pion optical potential, the scalar quark condensate $<{\bar q}q>$, the in-medium pion decay constant ${\tilde F}$, and the equation of state (EOS) of pion matter. 

The outline of the paper is as follows: In the next Sect., we find for the pion field a parameterization, which provides a constant Lagrange measure for path integrals. The integration over the pion field variables is extended from $- \infty$ to $+ \infty$ to convert path integrals to the Gaussian form. Such a procedure 
does apparently not affect the perturbative part of the Green functions. In Sect. III, a method for calculation of thermal averages within ChPT in the MF approximation is described. The thermal averages appear as inverse Borel transforms with respect to the non-renormalized scalar pion density. The lowest order expansion coefficients are calculated for the pion effective mass and other quantities and compared to the earlier calculations. In Sect. IV, we show that the power series with respect to the density are asymptotic and hence divergent. They can, however, be summed up with the help of the contour-improved Borel resummation technique. In Sect. V, predictions of the non-linear sigma model are compared to LTGs. The results obtained by the resummation are discussed in Conclusion.

\section{Pion field parameterization}
\setcounter{equation}{0} 

The lowest order ChPT Lagrangian has the form 
\begin{equation}
\mathcal{L}=\frac{F^{2}}{4}{\mbox Tr}[\partial _{\mu }U\partial ^{\mu
}U^{\dagger }]+\frac{F^{2}M_{\pi }^{2}}{4}{\mbox Tr}[U^{\dagger }+U]
\label{LAGR}
\end{equation}
where $M_{\pi }$ is the pion mass. The kinetic term for the matrix $U(x)$ $%
\in $ $SU(2)$ is invariant under the chiral transformations $U\rightarrow
U^{\prime }=RUL^{+}$ where $R,\,L$ $\in $ $SU(2)$. The second term in Eq.(%
\ref{LAGR}) breaks chiral symmetry explicitly. In what follows, we set $F=1$.

The matrix $U(x)$ can be parameterized in various ways. The on-shell vacuum
amplitudes do not depend on the choice of variables for pions \cite
{CHIK,KAME}. The method proposed by Gasser and Leutwyller \cite{GL84}
establishes a connection between QCD Green functions and amplitudes of the
effective chiral Lagrangian. Using this method, the QCD on- and off-shell
amplitudes can be calculated in a way independent on the parameterization.
The detailed studies of Refs. \cite{Tho95,Lee95} demonstrate that the
in-medium effective meson masses are independent on the choice of meson
field variables up to next-to-leading order in ChPT and to first order in
density, in accordance with the equivalence theorem. The in-medium off-shell
behavior and going beyond the linear-density approximation are discussed in
Refs. \cite{Lee95,Par02,Kon03}.

In order to provide the $S$-matrix invariant with respect to a symmetry
group, both, the action functional and the Lagrange measure entering the
path integral should be invariant. The chirally invariant measure $d\mu
[U]=d\mu [RUL^{+}]$ entering the path integral over the generalized
coordinates coincides with the Haar measure of the $SU(2)$ group. In the
exponential parameterization, 
\begin{equation}
U({\phi }^{\alpha })= e^{i{\tau }^{\alpha }{\phi }^{\alpha }}  \label{EP}
\end{equation}
and 
\begin{equation}
d\mu [U]=\sin ^{2}(\phi )\phi ^{-2}d^{3}\phi   \label{HAAR}
\end{equation}
(see e.g. \cite{WIGN}). 
The leading order ChPT is equivalent to the $O(4)$
non-linear sigma model due to the isomorphism of algebras $su(2)_{L}\oplus
su(2)_{R}\sim so(4)$. The quantization of the $O(4)$ non-linear sigma model 
\cite{MIK} yields automatically the measure (\ref{HAAR}).

Any perturbation theory uses for dynamical fields an oscillator basis to
convert path integrals into a Gaussian form. It is necessary to specify 
variables, $\chi ^{\alpha }$, in terms of which the Lagrange measure 
is ''flat'': 
\begin{equation}
d\mu [U]=d^{3}\chi .  \label{HAAR.CHI}
\end{equation}

The weight factor can always be exponentiated to generate an effective Lagrangian 
$\delta \mathcal{L}_{H}$, in which case $\chi ^{\alpha }=\phi ^{\alpha }$ provides 
the desired parameterization. The exponential parameterization (\ref{EP}) gives, 
in particular, $\delta \mathcal{L}_{H}= - \frac{1}{a^{4}}\log (\sin ^{2}(\phi )\phi ^{-2})$ 
where $a$ is a lattice size. $\delta \mathcal{L}_{H}$ diverges in the continuum limit. 
The non-linear sigma model is not a renormalizable theory, so divergences cannot be 
absorbed into a redefinition of $F$ and $M_{\pi }$. Using the MF approximation, it is 
usually possible to keep renormalizations finite. The exponentiation of a variable 
weight factor breaks, in general, selfconsistency of the MF approximation.

The divergences arising from $\delta \mathcal{L}_{H}$ could, however, be compensated by divergences coming from the higher orders ChPT loops. From this point of view it looks naturally to attribute $\delta \mathcal{L}_{H}$ to higher orders ChPT loop expansion starting from one loop. The MF approximation for the ChPT implies then that the tree level approximation neglects $\delta \mathcal{L}_{H}$ from the start. It is hard to expect that such an approximation is relevant at the high temperature limit where the chiral invariance is supposed to be restored.

The interaction terms in the effective Lagrangian which appear due to presence of the Haar measure have been discussed earlier in QCD \cite{WEISS,SAIL}. 
The exponentiation of the weight factor gives consistent results due to renormalizability of QCD. The divergences appearing in the continuum limit from $\delta \mathcal{L}_{H}$ at a tree level are compensated by one-loop gluon self-interaction diagrams.

The consistency of the MF approximation of the non-linear sigma model survives with one parameterization only, which uses the dilatated pion fields variables $\chi ^{\alpha }=\phi ^{\alpha }\chi /\phi $ such that $\chi ^{2}\chi
^{\prime }=\sin ^{2}(\phi )\geq 0$ where $\chi =(\chi ^{\alpha }\chi
^{\alpha })^{1/2}$ and
\begin{equation}
\chi ^{3}=\frac{3}{2}(\phi -\sin (\phi )\cos (\phi )).  \label{CHI}
\end{equation}
The vacuum value $\phi _{vac}^{\alpha }=0$ corresponds to $\chi_{vac}^{\alpha }=0$, $\chi $ is a monotonously increasing function of $\phi $. The value of $4\pi \chi ^{3}/3$ has the meaning of a volume covered by a $3$-dimensional surface of radius $\chi $ in a $4$-dimensional space.

The parameterization based on the dilatation of $\phi ^{\alpha }$ gives $\delta \mathcal{L}_{H}=0$, does not require the higher orders ChPT loop expansion for the consistency, and allows to work in the continuum limit with finite quantities only.

The $SU(2)$ group has a finite group volume. The magnitude of the $\chi ^{\alpha }$ fields is restricted by $\chi_{max}=(3\pi)^{1/3}$. It is generally believed that using the perturbation theory, one can extend the integrals over $\chi ^{\alpha }$ from $-\infty $ to $+\infty $. The modification of the result is connected to large fields fluctuations of a non-perturbative nature, which do, apparently, not affect perturbation series. This conjecture is essential for the standard loop expansion in ChPT. A similar extension of the integration region is used in QCD \cite{WEISS,SAIL}. 

The Weinberg's parameterization \cite{PWEIN} does not restrict the magnitude of the pion 
fields. Such a parameterization looks especially attractive as it simplifies conversion 
of the path integrals into the Gaussian form. Because of the variable Lagrange measure, 
it can be effective starting from one loop.

The propagators (heat kernels) of free particles moving on compact group manifolds have been analyzed in Refs.\cite{DOWK,MATE,SCHU,DURU,BAAQ}. It was shown that the semiclassical approximation for the propagators is exact. An explicit form of the propagator for the group $SU(2)$ is given by Schulman \cite{SCHU} and Duru \cite{DURU}. The problem of the variable Lagrange measure of the path integrals appears already at the quantum mechanical level. The path integral as it was recognized by Marinov and Terent'ev \cite{MATE} is ill posed at the tree level. As demonstrated by Baaquie \cite{BAAQ}, the divergent contributions coming from the variable Lagrange measure are cancelled by loops generated by the effective Lagrangian. A system of coupled oscillators on a compact group manifold represents unsolved problem. 

The requirement of chiral invariance of the Lagrange measure $d\mu[U]$ being combined with the requirements that (i) the perturbation expansion is based on the oscillator basis and (ii) the MF approximation does not involve loops restricts the parameterizations of the pion fields to only one admissible parameterization.

\section{Thermal averages in the MF approximation}
\setcounter{equation}{0} 

The quadratic part of the Lagrangian with respect to the fields $\chi
^{\alpha}$ can be extracted from Eq.(\ref{LAGR}) and the rest $\delta 
\mathcal{L}$ treated as a perturbation: 
\begin{equation}
\mathcal{L}=\frac{1}{2}\partial _{\mu }\chi ^{\alpha }\partial _{\mu }\chi
^{\alpha }-\frac{\tilde{M}_{\pi }^{2}}{2}\chi ^{2}+\delta \mathcal{L}
\label{LAGRANGIAN}
\end{equation}
where $\tilde{M}_{\pi }$ is an effective pion mass to be determined
self-consistently and 
\begin{equation}
\delta \mathcal{L}=\frac{1}{2}\partial _{\mu }\chi ^{\alpha } \partial _{\mu
} \chi ^{\alpha } \mathcal{J}_{0}+\frac{1}{2}\partial _{\mu }\chi ^{\alpha
}\partial _{\mu }\chi ^{\beta }\chi ^{\alpha }\chi ^{\beta }\chi ^{-2}%
\mathcal{J}_{1}+ M_{\pi }^{2}(\mathcal{J}_{2} - 1)+\frac{\tilde{M}_{\pi }^{2}%
}{2}\mathcal{J}_{3},  \label{DLAGR}
\end{equation}
with 
\begin{eqnarray}
\mathcal{J}_{0} &=&\frac{\sin ^{2}(\phi )}{\chi ^{2}} - 1,  \label{J0} \\
\mathcal{J}_{1} &=&\frac{\chi ^4}{\sin ^{4}(\phi )}-\frac{\sin ^{2}(\phi )}{%
\chi ^{2}},  \label{J1} \\
\mathcal{J}_{2} &=&\cos (\phi ),  \label{J2} \\
\mathcal{J}_{3} &=&\chi ^{2}.  \label{J3}
\end{eqnarray}

The physical observables are expressed in terms of the thermal Green functions.

\subsection{Thermal Green functions}

The scalar pion density $\rho $ in a thermal bath with temperature $T$ is
defined by 
\begin{equation}
\delta ^{\alpha \beta }2\rho =<{\chi }^{\alpha }{\chi }^{\beta }>=\delta
^{\alpha \beta }Z_{\chi }\int {\ \frac{d{}\mathbf{k}}{(2\pi )^{3}}2}n_{0}(%
\mathbf{k})  \label{PDEN}
\end{equation}
where $Z_{\chi }$ is a renormalization constant of the pion $\chi $-field
and 
\begin{equation}
n_{0}(\mathbf{k})={\frac{1}{2\omega ({}\mathbf{k})}\frac{1}{e^{\frac{\omega (%
\mathbf{k}{})}{T}}-1}.}  \label{PSDEN}
\end{equation}
The average (\ref{PDEN}) involves the normally ordered $\chi^{\alpha}$ operators. 
The energy $\omega (\mathbf{k})=\sqrt{\tilde{M}_{\pi }^{2}+{}\mathbf{k}^{2}%
}$ which enters the Bose-Einstein distribution contains the in-medium pion
mass ${\tilde{M}}_{\pi }$. A similar situation occurs in the Bogoliubov
model of a weakly interacting non-ideal Bose gas. Also in Fermi liquid
theory the momentum space distribution is determined by the in-medium
dispersion law of quasi-particles (see e.g. \cite{ABRI}).

In terms of thermal two-body Green function,
\begin{equation}
\mathcal{G}^{\alpha \beta }(\tau ,\mathbf{x})=-<\mathcal{T}{\chi }^{\alpha }(\tau ,
\mathbf{x}){\chi }^{\beta }(0,\mathbf{0})>,  \label{GF}
\end{equation}
the momentum space distribution $n_{0}(\mathbf{k})$ is determined by equation
\begin{equation}
\delta ^{\alpha \beta }Z_{\chi } (2n_{0}(\mathbf{k}) + \frac{1}{2\omega(\bf{k})})=-T\lim_{\epsilon \rightarrow
-0}\sum_{s=-\infty }^{+\infty }\mathcal{G}^{\alpha \beta }(\omega _{s},\mathbf{k}%
)e^{-i\omega _{s}\epsilon }  \label{MSDIS}
\end{equation}
where $\omega _{s}=2\pi sT$ and 
\begin{equation}
\mathcal{G}^{\alpha \beta }(\tau ,\mathbf{x})=T\sum_{s=-\infty }^{+\infty }\int {\ 
\frac{d{}\mathbf{k}}{(2\pi )^{3}}}\mathcal{G}^{\alpha \beta }(\omega _{s},\mathbf{k}%
)e^{-i\omega _{s}\tau +i\mathbf{kx}}.  \label{FTGF}
\end{equation}

The density $\rho $ can be found from
\begin{equation}
\delta ^{\alpha \beta }(2\rho + \rho_{vac}) =-\lim_{\epsilon \rightarrow -0}\mathcal{G}^{\alpha \beta
}(\epsilon ,\mathbf{0}),  \label{GFDEN}
\end{equation}
where $\rho_{vac}$ is the vacuum density of the zero-point field fluctuations
\[
\rho_{vac} = Z_{\chi }\int {\ \frac{d{}\mathbf{k}}{2\omega({\mathbf k})(2\pi )^{3}}}.
\]
The vacuum density can be treated self-consistently beyond the lowest order ChPT only. 
In the MF approximation it is assumed that zero-point field fluctuations are absorbed to a redefinition of physical parameters entering the Lagrangian. In our case, the pion mass and the pion decay constant receive these contributions.

We thus neglect the vacuum fluctuations. In what follows, $\rho_{vac}$ is set equal to zero wherever it appears. This is equivalent to using thermal part of the Green function within the loops:
\begin{equation}
  \lim_{\epsilon \rightarrow -0} \mathcal{G}^{\alpha \beta }(\epsilon ,\mathbf{0}) 
  \rightarrow 
  \lim_{\epsilon \rightarrow -0} \mathcal{G}^{\alpha \beta }(\epsilon ,\mathbf{0})
- \lim_{\epsilon \rightarrow -0} \mathcal{G}^{\alpha \beta }(\epsilon ,\mathbf{0})|_{T=0}.
\label{GFTE}
\end{equation}

\begin{figure}[!htb]
\begin{center}
\includegraphics[angle=0,width=1cm]{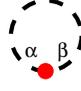}
\caption{
Diagram representation of the thermal average $<{\chi }^{\alpha }{\chi }^{\beta }>$ 
defined by Eqs.(\ref{PDEN}) and (\ref{GFDEN}).
The thick dashed line represents the dressed pion propagator.}
\label{fig1}
\end{center}
\end{figure}

Eq.(\ref{GFDEN}) is illustrated graphically on Fig. \ref{fig1}. Since we neglect
the zero-point field fluctuations,
we may not worry on the ordering of operators, entering thermal averages such 
as (\ref{PDEN}), (\ref{SIGMA}) and others, anymore.


\subsection{In-medium pion mass and other observables}

\begin{figure}[!htb]
\begin{center}
\includegraphics[angle=0,width=4cm]{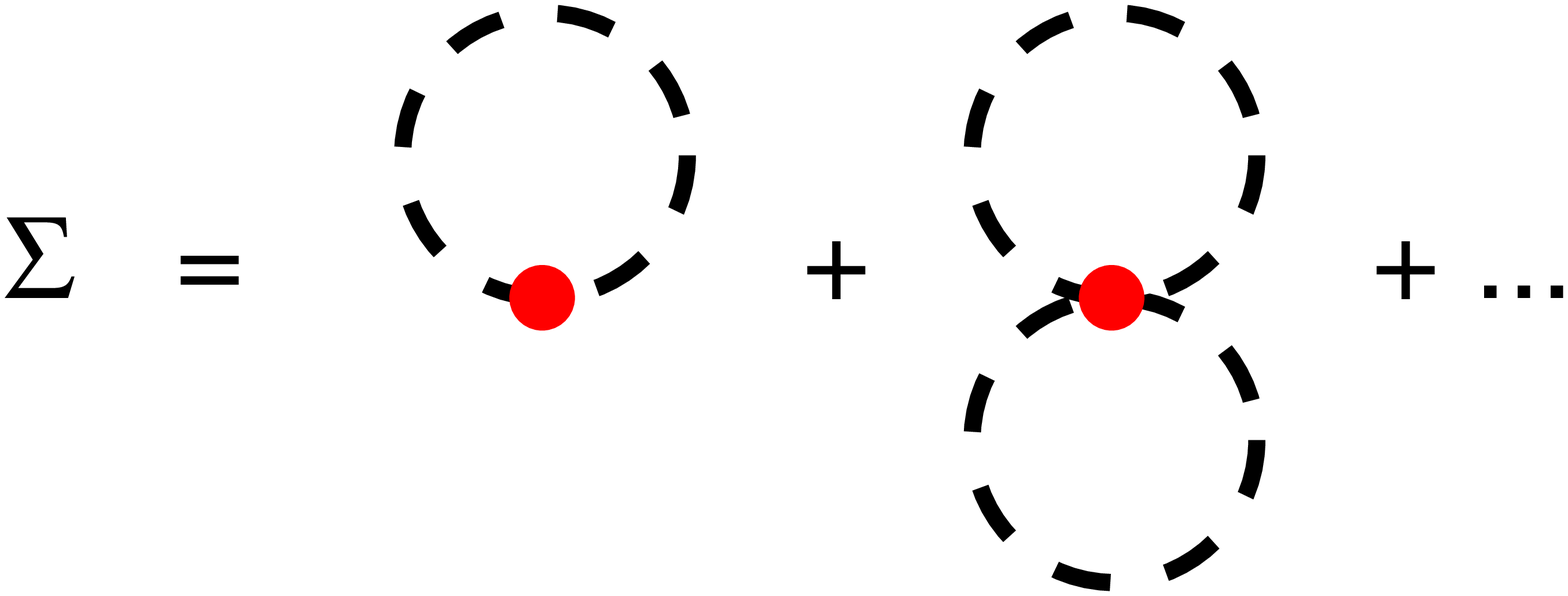}
\caption{
Diagram representation of the self-energy operator in the MF approximation. 
}
\label{fig3}
\end{center}
\end{figure}

In the MF approximation, the self-energy operator can be calculated from 
\begin{equation}
\delta ^{\alpha \beta }\Sigma (x-y)=-<\frac{\delta ^{2}}{\delta {\chi }%
^{\alpha }(x)\delta {\chi }^{\beta }(y)}\int \delta \mathcal{L}(z)d^{4}z>.
\label{SIGMA}
\end{equation}
Equation $<{\chi }^{\alpha }\partial _{\mu }{\chi }^{\beta }>=0$ is a
consequence of the symmetry of the pion matter with respect to isospin
rotations. In the momentum representation, using equation 
\begin{equation}
<\partial _{\mu }{\chi }^{\alpha }\partial _{\mu }{\chi }^{\beta }>=\delta
^{\alpha \beta }{\tilde{M}}_{\pi }^{2}2\rho ,  \label{LINE}
\end{equation}
one gets 
\begin{equation}
\Sigma (k^{2})=k^{2}\Sigma _{1}+M_{\pi }^{2}\Sigma _{2}+\tilde{M}_{\pi
}^{2}\Sigma _{3}  \label{self}
\end{equation}
where $k^{2}=-\omega _{s}^{2}-\mathbf{k}^{2}$ and 
\begin{eqnarray}
\Sigma _{1} &=&-<\mathcal{J}_{0}+\frac{1}{3}\mathcal{J}_{1}>,  \label{SIGMA1}
\\
\Sigma _{2} &=&\frac{1}{2\rho }<(1-\frac{\chi ^{2}}{6\rho })\mathcal{J}_{2}>,
\label{SIGMA2} \\
\Sigma _{3} &=&\frac{3}{2}<(1-\frac{\chi ^{2}}{6\rho })(\mathcal{J}_{0}+%
\frac{1}{3}\mathcal{J}_{1})>-1\text{.}  \label{SIGMA3}
\end{eqnarray}
The self-energy operator is presented graphically on Fig. \ref{fig3}

Eq.(\ref{LINE}) can be derived as follows: The thermal average (\ref{LINE}) 
can be rewritten in terms of the two-body Green function
\begin{equation}
<\partial _{\mu }\chi ^{\alpha }\partial _{\mu }\chi ^{\beta }>= 
\lim_{\tau \rightarrow +0}\Box \mathcal{G}^{\alpha \beta }(\tau ,\mathbf{0}%
)  \label{B1}
\end{equation}
where $\Box =-\frac{\partial ^{2}}{\partial \tau ^{2}}-\Delta $. The
two-body Green function satisfies equation
\begin{equation}
(\Box +\tilde{M}_{\pi }^{2}+\Sigma (-\Box ))\mathcal{G}^{\alpha \beta }(\tau
,\mathbf{x})=-\delta ^{\alpha \beta }\delta (\tau )\delta ^{3}(\mathbf{x})..
\label{B2}
\end{equation}
The perturbation $\delta \mathcal{L}(z)$ is quadratic with respect to the
derivatives, so $\Sigma (k^{2})$ in the MF approximation is a first order
polynomial. The expansion (\ref{self}) makes this feature explicit. Eq.(\ref
{B2}) can be rewritten in the form
\begin{equation}
(\Box +\tilde{M}_{\pi }^{2})\mathcal{G}^{\alpha \beta }(\tau ,\mathbf{x}%
)=-Z_{\chi }\delta ^{\alpha \beta }\delta (\tau )\delta ^{3}(\mathbf{x})
\label{B9}
\end{equation}
where $Z_{\chi }$ is the renormalization constant introduced earlier. Using
$\lim_{\tau \rightarrow +0}\delta (\tau )=0$ and Eqs.(\ref{GFDEN}), (\ref{B1}), 
(\ref{B9}), and (\ref{UNIV}), we arrive at Eq.(\ref{LINE}) with $2\rho$ replaced 
by $2\rho + \rho_{vac}$. The value of $\rho_{vac}$ must be neglected, however.

\begin{figure}[!htb]
\begin{center}
\includegraphics[angle=0,width=6.3cm]{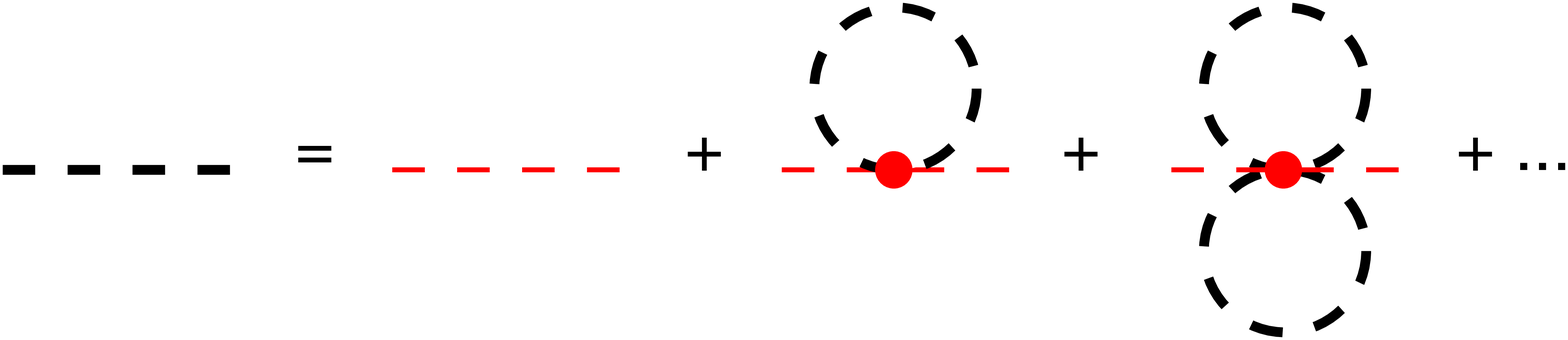}
\caption{
Diagram representation of the dressed pion propagator (thick dashed lines)
in terms of the scalar density and the bare pion 
propagators (thin dashed lines). The scalar density appears from loops formed by 
the dressed pion propagator according to Eq.(\ref{GFDEN}).
}
\label{fig5}
\end{center}
\end{figure}

The effective pion mass can be determined from equation $\Sigma (\tilde{M}%
_{\pi }^{2})=0$: 
\begin{equation}
\tilde{M}_{\pi }^{2}=-M_{\pi }^{2}\frac{\Sigma _{2}}{\Sigma _{1}+\Sigma _{3}}%
,  \label{EMAS}
\end{equation}
while the renormalization constant is given by 
\begin{equation}
Z_{\chi }^{-1}=1-\frac{\partial \Sigma (k^{2})}{\partial k^{2}}=1-\Sigma
_{1}.  \label{RENO}
\end{equation}
The Green function in the MF approximation has the form 
\begin{equation}
\mathcal{G}^{\alpha \beta }(\omega _{s},\mathbf{k})=-\delta ^{\alpha \beta }\frac{%
Z_{\chi }}{\omega _{s}^{2}+\mathbf{k}^{2}+{\tilde{M}}_{\pi }^{2}}.
\label{GFMF}
\end{equation}
It is presented graphically on Fig. \ref{fig5}.

In what follows, we need also functions 
\begin{eqnarray}
\mathcal{J}_{4} &=&\frac{\sin (\phi )}{\chi },  \label{J4} \\
\mathcal{J}_{5} &=&-\mathcal{J}_{2}\left( -\mathcal{J}_{4}+(20\rho -2\chi
^{2})\mathcal{J}_{4}^{\prime }+8\rho \chi ^{2}\mathcal{J}_{4}^{\prime \prime
}\right)  \nonumber \\
&&+\mathcal{J}_{4}\left( (12\rho -2\chi ^{2})\mathcal{J}_{2}^{\prime }+8\rho
\chi ^{2}\mathcal{J}_{2}^{\prime \prime }\right)  \label{J5}
\end{eqnarray}
which appear in calculations of density dependent renormalization 
constant $Z_{\pi }$ and the pion decay constant $\tilde{F}$ according to 
Eqs. (\ref{SIGMA4}) and (\ref{SIGMA6}). Here, $\mathcal{J}_{i}^{\prime}$ 
and $\mathcal{J}_{i}^{\prime \prime }$ are first and second derivatives 
with respect to $\phi$.

\begin{figure}[!htb]
\begin{center}
\includegraphics[angle=0,width=7.53cm]{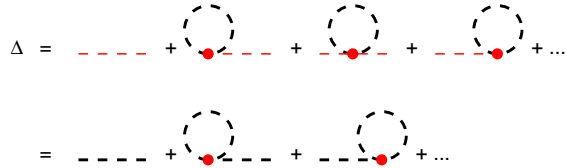}
\caption{
Diagram representation of the pion propagator $\Delta^{\alpha \beta }(x-y)$. 
The dressed pion propagators are shown as thick dashed lines and the bare pion 
propagators are shown as thin dashed lines.
The second line
contains insertions into the ends of the dressed pion propagators only.
}
\label{fig4}
\end{center}
\end{figure}

The thermal Green function of the pion is defined by 
$\Delta^{\alpha \beta }(x-y)=-<\mathcal{T}\pi ^{\alpha }(x)\pi ^{\beta }(y)>$. The pion
fields $\pi ^{\alpha }(x)=$ $\frac{1}{2}{\mbox Tr}[\tau ^{\alpha }U(x)]$ do
not depend on derivatives of $\chi ^{\alpha }$. In such a case, no
additional $k^{2}$ dependence appears as compared to the $\chi $ propagator.
Given the renormalization constant $Z_{\chi }$, the renormalization constant 
$Z_{\pi }$ of the pion propagator can be found from 
\begin{equation}
\delta ^{\alpha \beta }Z_{\pi }^{1/2}Z_{\chi }^{-1/2}=<\frac{\partial \pi
^{\alpha }}{\partial \chi ^{\beta }}>.  \label{ZPIZICHI}
\end{equation}
The pion propagator $\Delta^{\alpha \beta }(x-y)$ is shown graphically on Fig. \ref{fig4}.

The quark scalar condensate $<\bar{q}q>$ is proportional to the expectation
value of the scalar field $\sigma =\frac{1}{2}{\mbox Tr}[U(x)]$. The calculation
of the averages $<{\partial \pi^{\alpha }}/{\partial \chi ^{\beta }}>$ and 
$<\sigma>$ leads after factorization of isotopic indices to the calculation of averages defined in terms of 
functions (\ref{J2}) and (\ref{J4}):
\begin{eqnarray}
Z_{\pi }^{1/2}Z_{\chi }^{-1/2} &=& < \frac{\chi ^2}{6\rho} \mathcal{J}_{4}>,
\label{SIGMA4} \\
<\bar{q}q>/<\bar{q}q>_{vac} &=&<\mathcal{J}_{2}>.  \label{SIGMA5}
\end{eqnarray}

\begin{figure}[!htb]
\begin{center}
\includegraphics[angle=0,width=4.6cm]{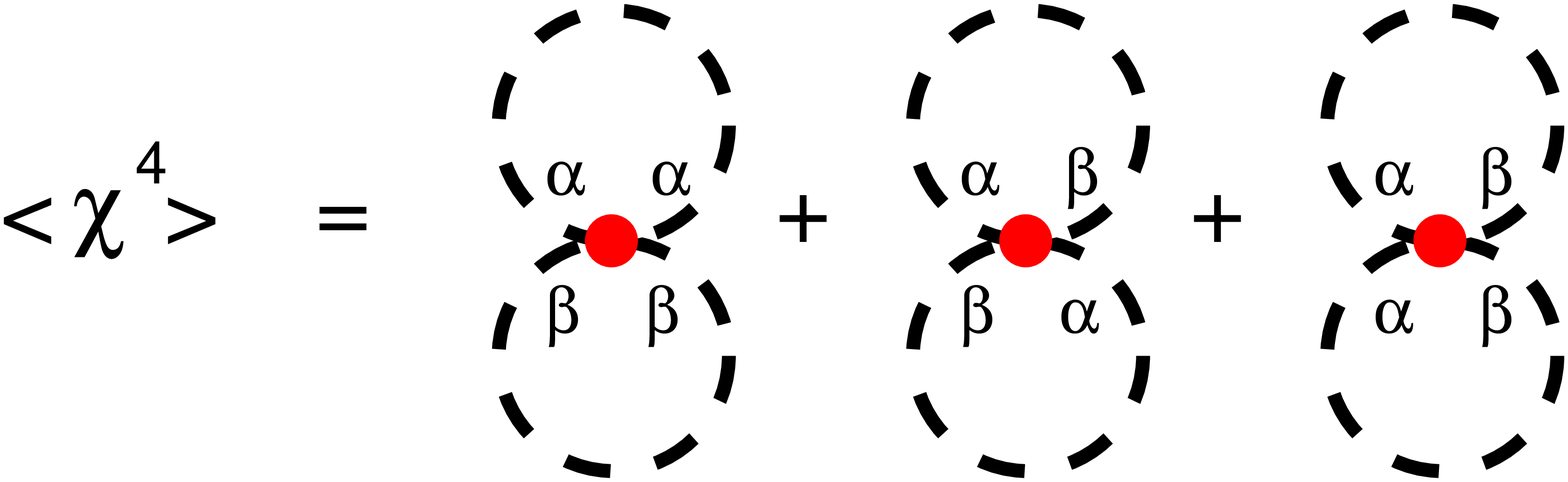}
\caption{
Diagram representation of the thermal average $<\chi ^{4}>=<\chi ^{\alpha }\chi ^{\alpha }\chi ^{\beta }\chi ^{\beta }> = 
(3\times 3+3+3)\times 4\rho ^{2}$. 
Greek indices e.g. $\alpha$ and $\beta$ within one loop designate $\delta^{\alpha \beta}$. The result $60\rho ^{2}$ is in agreement with Eq.(\ref{CALC}) for $n=2$. 
}
\label{fig2}
\end{center}
\end{figure}

In the pion gas, the spectral density of the axial two-point function $%
<\mathcal{T}A_{\mu }^{\alpha }(x)A_{\nu }^{\beta }(y)>$ is characterized by two
different pion decay constants \cite{TOUB97}. In the limit $T=0$, these
constants coincide with $F$. Let us find an in-medium pion decay constant, ${%
\tilde{F}}$, appearing in the spectral density of the two-point function $%
<\mathcal{T}\partial _{\mu }A_{\mu }^{\alpha }(x)\partial _{\nu }A_{\nu }^{\beta }(y)>$%
.. This definition is equivalent to equation $\partial _{\mu }A_{\mu
}^{\alpha }=\tilde{F}\tilde{M}_{\pi }^{2}Z_{\pi }^{-1/2}\pi ^{\alpha }$
where the right side gives the physical pion $Z_{\pi }^{-1/2}\pi ^{\alpha }$%
.. The axial current has the form $A_{\mu }^{\alpha }=-\sigma \partial _{\mu
}\pi ^{\alpha }+(\partial _{\mu }\sigma )\pi ^{\alpha }$. 
The calculation of the average $<\mathcal{T}A_{\mu }^{\alpha }(x)A_{\nu }^{\beta }(y)>$
leads after factorization of isotopic indices to the calculation of an average defined in terms of the
function (\ref{J5}). In this way we obtain
\begin{equation}
\tilde{F}Z_{\chi }^{-1/2}=<\frac{\chi ^{2}}{6\rho }\mathcal{J}_{5}>. 
\label{SIGMA6}
\end{equation}

\subsection{Thermal averages as inverse Borel transforms}

Suppose we want to find the thermal average of a function $\mathcal{J}({\chi
^{2}})$ which can be expanded in a power series of ${\chi }^{2}$. The
average values of powers of the operator ${\chi }^{2}={\chi }^{\gamma }{\chi 
}^{\gamma }$ can be calculated as follows: 
\begin{eqnarray}
<({\chi }^{\gamma }{\chi }^{\gamma })^{n}> &=&<\sum_{s_{1}s_{2}s_{3}}\frac{n!%
}{s_{1}!s_{2}!s_{3}!}({\chi }^{1})^{2s_{1}}({\chi }^{2})^{2s_{2}}({\chi }%
^{3})^{2s_{3}}>  \nonumber \\
&=&\sum_{s_{1}s_{2}s_{3}}\frac{n!}{s_{1}!s_{2}!s_{3}!}<({\chi }%
^{1})^{2s_{1}}><({\chi }^{2})^{2s_{2}}><({\chi }^{3})^{2s_{3}}>  \nonumber \\
&=&\sum_{s_{1}s_{2}s_{3}}\frac{n!}{s_{1}!s_{2}!s_{3}!}%
(2s_{1}-1)!!(2s_{2}-1)!!(2s_{3}-1)!!(2\rho )^{n}  \nonumber \\
&=&\sum_{s_{1}s_{2}s_{3}}\frac{n!}{s_{1}!s_{2}!s_{3}!}\frac{1}{\pi ^{3/2}}%
\Gamma (s_{1}+\frac{1}{2})\Gamma (s_{2}+\frac{1}{2})\Gamma (s_{3}+\frac{1}{2}%
)(4\rho )^{n}  \nonumber \\
&=&\sum_{s_{1}s_{2}s_{3}}\frac{n!}{s_{1}!s_{2}!s_{3}!}\frac{1}{\pi ^{3/2}}%
(4\rho )^{n}\int_{0}^{\infty }\int_{0}^{\infty }\int_{0}^{\infty
}x_{1}^{s_{1}-\frac{1}{2}}x_{2}^{s_{2}-\frac{1}{2}}x_{3}^{s_{3}-\frac{1}{2}%
}e^{-(x_{1}+x_{2}+x_{3})}dx_{1}dx_{2}dx_{3}  \nonumber \\
&=&\frac{1}{\pi ^{3/2}}(4\rho )^{n}\int_{0}^{\infty }\int_{0}^{\infty
}\int_{0}^{\infty }\frac{(x_{1}+x_{2}+x_{3})^{n}}{\sqrt{x_{1}x_{2}x_{3}}}%
e^{-(x_{1}+x_{2}+x_{3})}dx_{1}dx_{2}dx_{3}  \nonumber \\
&=&\frac{4}{\sqrt{\pi }}\int_{0}^{\infty }{\ (4\rho
x^{2})^{n}x^{2}e^{-x^{2}}dx}  \nonumber \\
&=&\frac{2}{\sqrt{\pi }}\int_{0}^{\infty }{\ (4\rho t)^{n}t^{1/2}e^{-t}dt}..
\label{CALC}
\end{eqnarray}

The summations run for $s_{1}+s_{2}+s_{3}=n$, where $s_{i}=0,1,...$. In the
first line, we use the multibinomial formula for $(({\chi }^{1})^{2}+({\chi }%
^{2})^{2}+({\chi }^{3})^{2})^{n}$. The factorization of the average in the second line
is a consequence of the MF approximation according to which the averages are calculated over
the free gas of quasiparticles, i.e., non-interacting effective pions in three isotopic states.
In order to calculate, e.g., the average
value $<({\chi }^{1})^{2s_{1}}>$, one has to consider $(2s_{1})!$
permutations of the operators ${\chi }^{1}$ due to their possible pairings.
Permutations inside of each pair are counted twice, whereas permutations of
the pairs are counted $s_{1}!$ times. This gives the factor $%
(2s_{1})!/(2^{s_{1}}s_{1}!)=(2s_{1}-1)!!$ in the third line of (\ref{CALC}%
). To arrive at the final result one has to express $(2s_{i}-1)!!$ in terms
of Euler's gamma functions and use their integral representations to convert
tree-dimensional integral into the one-dimensional integral. 

The MF approximation we used consists in the neglection of dynamical loops, i.e., loops 
composed from more than one dressed pion propagator. The loops composed from one
dressed pion propagator give the density (\ref{PDEN}) after neglecting $\rho_{vac}$. 
The calculation of the $n=1$ average value is graphically presented on Fig.. \ref{fig1}
(the indices $\alpha $ and $\beta $ have to be contracted with $\delta
^{\alpha \beta }$). The calculation of the $n=2$ is illustrated on Fig. \ref
{fig2}. An alternative proof of Eq.(\ref{CALC}), based on the thermal Green 
functions method, is given in Appendix A.

The average value of $\mathcal{J}({\chi ^{2}})$ can be written in the form 
\begin{equation}
<\mathcal{J}({\chi ^{2}})>=\frac{2}{\sqrt{\pi }}\int_{0}^{\infty }{\mathcal{J%
}(4\rho t)t^{1/2}e^{-t}dt}.  \label{MAIN}
\end{equation}
The thermal fluctuations are suppressed exponentially according to the value of 
$\chi ^{2}$ (see also Eq.(\ref{MAINP})). For small fluctuations, $\chi $ coincides with $\phi $.

The thermal average $<\mathcal{J}(\chi^2)>$ to all orders in the
pion density and to leading order in ChPT is given by the inverse Borel
transform of the function $\frac{2}{\sqrt{\pi}}\mathcal{J}(4t)$.

Using the explicit form of $\mathcal{J}_{i} $ and re-expanding results in
terms of the physical density $\rho _{0}=Z_{\chi }^{-1}\rho$, we obtain for $%
\rho_0 << 1$ 
\begin{eqnarray}
\tilde{M}_{\pi }^{2}/M_{\pi }^{2} &=&1+\rho _{0}-\frac{39}{10}\rho _{0}^{2} -%
\frac{2601}{50}\rho _{0}^{3}-\frac{614163}{1400}\rho _{0}^{4}+...,
\label{M2EFF} \\
Z_{\chi }^{-1} &=&1+\frac{12}{5}\rho _{0}^{2}+\frac{368}{25}\rho _{0}^{3}+%
\frac{80832}{875}\rho _{0}^{4}+...,  \label{ZICHI} \\
Z_{\pi }^{-1} &=&1+2\rho _{0}+\frac{46}{5}\rho _{0}^{2}+\frac{1108}{25}\rho
_{0}^{3}+\frac{204202}{875}\rho _{0}^{4}+...,  \label{ZIPIO} \\
<\bar{q}q>/<\bar{q}q>_{vac} &=&1-3\rho _{0}-\frac{3}{2}\rho _{0}^{2}+\frac{39%
}{10}\rho _{0}^{3}+\frac{7803}{200}\rho _{0}^{4}+...,  \label{SCALA} \\
\tilde{F}/F &=&1-2\rho _{0}+\frac{2}{5}\rho _{0}^{2}+\frac{156}{25}\rho
_{0}^{3}+\frac{22674}{875}\rho _{0}^{4}+... \; .  \label{FTILD}
\end{eqnarray}

For massless pions $\rho _{0}=\frac{1}{24}T^{2}$. In such a case, the power
series expansions over density and temperature coincide.

The result for $\tilde{M}_{\pi }^{2}$ to order $O(\rho _{0})$ is in
agreement with \cite{GL87,TOUB97,KRIV}, the result for $Z_{\pi }^{-1}$ to
order $O(\rho _{0})$ is in agreement with \cite{KRIV}, the result for $<\bar{%
q}q>$ to order $O(\rho _{0}^{2})$ is in agreement with \cite
{GL87,GELE89,ELET94,BOKA96,TOUB97}, the result for $\tilde{F}$ to order $%
O(\rho _{0})$ is in agreement with \cite{GL87,ELET93,BOKA96,JEON}.

\begin{figure}[!htb]
\vspace{20mm}
\begin{center}
\includegraphics[width=.56\textwidth,height=.394\textheight]{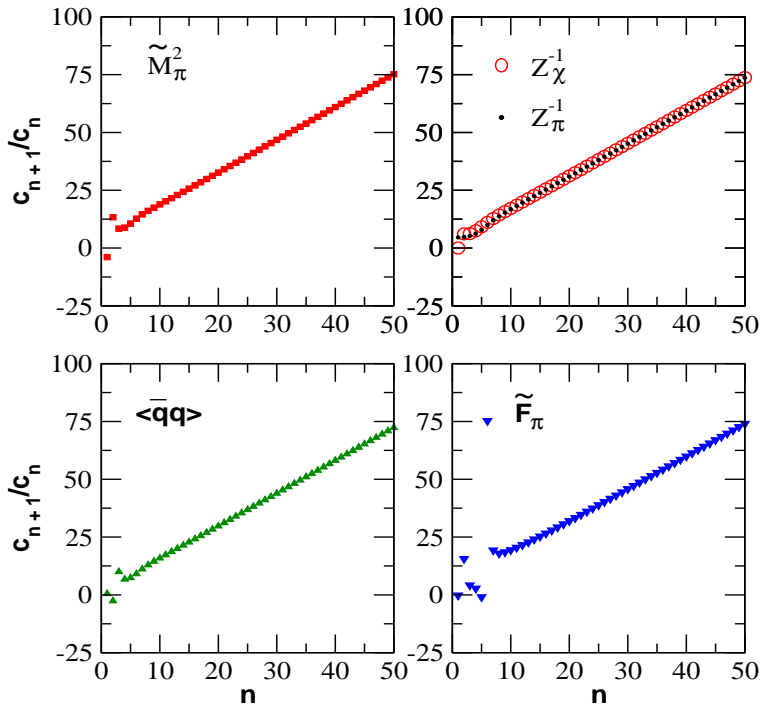}
\caption{
The ratios $c_{n + 1}/c_{n}$ of the expansion coefficients of power series
$\sum_{k=0}^{\infty}c_{n}\rho_0^{n}$ over the pion density $\rho_0$ 
for (a) the in-medium pion mass ${\tilde M}_{\pi}^2$, 
(b) the propagator renormalization constants $Z_{\chi}^{-1}$ of the $\chi$-field 
and $Z_{\pi}^{-1}$ of the pion field,
(c) the scalar quark condensate $<{\bar q}q>$ , and (d) the 
pion decay constant ${\tilde F}$.
The linear growth of the ratios beyond $n \sim 10$ indicates that all power series
are asymptotic. The same sign of $c_n$ for $n > 10$
indicates furthermore that the series are not Borel summable. 
The approximately equal slopes are related to a singularity of the Borel
transforms on the real axis at $\phi = \pi$ (see text). 
}
\label{levels}
\end{center}
\end{figure}

Chiral invariance of the non-linear sigma model in the MF 
approximation is commented in Appendix B.

\section{Summation of the density series}
\setcounter{equation}{0} 

Most perturbation series in quantum field theory are believed to be
asymptotic and hence divergent \cite{DYSON,ITZU}. The integral
representation (\ref{MAIN}) permits by changing the variable $t \rightarrow
t^{\prime} = t/\rho$ an analytical continuation into the complex half-plane $%
{e}[\rho] > 0$, whereas at ${e}[\rho] < 0$ the integral diverges. The
half-plane ${e}[\rho] < 0$ contains in general singularities the character
of which depends on properties of the function $\mathcal{J}({\chi ^2})$. The
convergence radius of the Taylor expansion is determined by the nearest
singularity. One cannot exclude that the point $\rho = 0$ around which we
make the expansion is a singular point.

Inspecting Eqs.(\ref{J0})-(\ref{J3}) and (\ref{J4})-(\ref{J5}), we observe
that the Borel transforms involving $\mathcal{J}_{1}$ and $\mathcal{J}_5$
are singular at $\phi = k\pi$ where $k=1,2,...$. The integral along the real
half-axis of the Borel variable $t$ is therefore not feasible. Series
involving such functions are asymptotic, have zero convergence radii, and
are furthermore not Borel summable \cite{ITZU,BEND,JEAN}.

In order to clarify the character of the power series over the physical
density $\rho_0$, we calculate the higher order expansion coefficients. We
write ${O}(\rho_0) = \sum_{k=0}^{\infty}c_{n}\rho_0^{n}$ for observables (\ref
{M2EFF})-(\ref{FTILD}) and plot on Fig. \ref{levels} the ratios $c_{n +
1}/c_{n}$ versus $n$ up to $n = 50$. \footnote{%
The MAPLE code used to calculate the expansion coefficients is availbale
upon request.} The apparent asymptotic regime starts at $n \sim 10$. The
ratios $c_{n + 1}/c_{n}$ increase linearly with $n$, indicating clearly that
the power series are asymptotic. The slopes of these ratios are
approximately equal. One can expect that the nearest singularity $\phi = \pi$
that appears in the series expansion over $\rho$ appears again in the series
expansion over $\rho_0$ by virtue of $\rho = Z_{\chi }\rho _{0}$. In such a
case, we would expect a singularity in the Borel plane at $4t=(\frac{3\pi}{2}%
)^{2/3}$ as well. From the other side, for $c_{n} \sim n!a^{n}$ the Borel
transform is singular at $t = 1/a$. The slope equals then $a=(\frac{16}{3\pi}%
)^{2/3} \simeq 1.4$, in good agreement with results presented on Fig. \ref
{levels}. The power series (\ref{M2EFF})-(\ref{FTILD}) with respect to the
physical density $\rho_{0}$ have zero convergence radii and are also not
Borel summable.

Generalizations of the Borel summation method have been proposed and
effectively used in mathematics and atomic physics \cite{BEND,JEAN}. In our
case, fortunately, the residues at $\phi =k\pi $ of the Borel transforms
vanish. A contour-improved Borel resummation technique can therefore be
applied which consists in a shift of the integration contour into the
complex $t$-plane. The result is stable against small variations of the
contour around the positive real axis. The thermal averages, in particular,
do not acquire imaginary parts and thus the expectation value, e.g., of the
hermitian operator ${\bar{q}}q$ remains real. The uniqueness of the result
justifies the integrations by parts of the functions ${}\mathcal{J}_{i}$,
performed to derive Eqs.(\ref{SIGMA2}), (\ref{SIGMA3}), (\ref{SIGMA4}), and 
(\ref{SIGMA6}).

\begin{figure}[!htb]
\vspace{5mm}
\begin{center}
\includegraphics[width=.3236\textheight,height=.20\textheight]{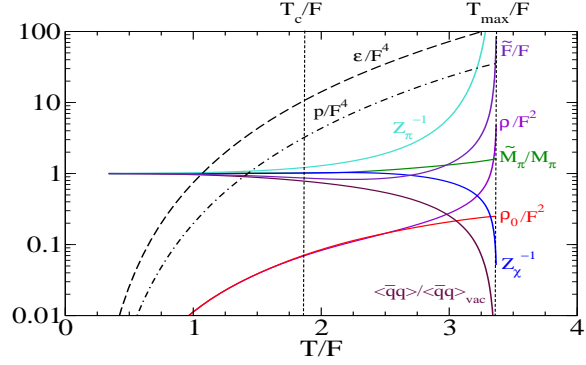}
\caption{
The non-renormalized scalar pion density $\rho$, the renormalized scalar pion density $\rho_0$,
the ratio ${\tilde M}_{\pi}/M_{\pi}$ between the in-medium and the vacuum pion mass,
the inverse renormalization constants $Z^{-1}_{\chi}$ and $Z^{-1}_{\pi}$ of the pion $\chi$-field 
and the pion field $\pi ^{\alpha }(x)=$ $\frac{1}{2}{\mbox Tr}[\tau ^{\alpha }U(x)]$, respectively,
the scalar quark condensate $<{\bar q}q>$, and the in-medium 
pion decay constant ${\tilde F}$ versus temperature $T$ are shown. 
The energy density $\varepsilon$
(dashed curve) and the pressure $p$ (dot-dashed curve) of the pion gas are also 
plotted. Solutions do not exist above $T_{max} = 3.36F \simeq 310$ MeV (the right vertical line). 
The critical temperature $T_{c} \simeq 175$ MeV of phase transition into the quark matter 
is shown by the left vertical line. All quantities are given 
in units of the pion decay constant $F=93$ MeV. 
}
\label{dendep}
\end{center}
\end{figure}

The power series (\ref{M2EFF})-(\ref{FTILD}) become summable using
the contour-improved Borel resummation method.

The EOS of pion matter can be determined by averaging the energy-momentum
tensor constructed from the Lagrangian (\ref{LAGRANGIAN}). The renormalized
energy density $\varepsilon =Z_{\chi }^{-1}T_{00}$ and the pressure $p=Z_{\chi
}^{-1}\sum_{i=1,3}T_{ii}/3$ are given by 
\begin{eqnarray}
\varepsilon &=&6\int {\ \frac{d{}\mathbf{k}}{(2\pi )^{3}}}\omega ^{2}({}\mathbf{%
k})n_{0}(\mathbf{k})-Z_{\chi }^{-1}\left( \tilde{M}_{\pi }^{2}3\rho
_{0}+M_{\pi }^{2}(\sigma -1)\right) ,  \label{EDEN} \\
p &=&2\int {\ \frac{d{}\mathbf{k}}{(2\pi )^{3}}}\mathbf{k}^{2}n_{0}(\mathbf{k%
})+Z_{\chi }^{-1}\left( \tilde{M}_{\pi }^{2}3\rho _{0}+M_{\pi }^{2}(\sigma
-1)\right)  \label{PRES}
\end{eqnarray}
where $\sigma =<\bar{q}q>/<\bar{q}q>_{vac}$. The results are plotted in Fig. 
\ref{dendep}.

There exists phenomenologically a broad interval of $0 < T < 2.48F \simeq 230$ MeV with $Z_{\chi }<1$%
, covering the range of temperatures specific for heavy-ion colliders. Above 
$T\simeq 230$ MeV the renormalization constant $Z_{\chi }^{-1}<1$. In the
vacuum, one has $Z^{-1}>1$ as a consequence of positive definiteness of
the spectral density of two-point Green functions. In the medium, the
spectral density of temperature Green functions of bosons is not
positive definite (see e.g. \cite{ABRI}, Chap. 17-3), so $Z_{\chi }^{-1}<1$
does apparently not contradict unitarity. Although the product of $%
\omega $ and the pion spectral density cannot be negative, sum rule for
this product is unknown. The region of
self-consistency of the MF approximation within perturbation theory over the density
extends up to $T_{max} = 3.36F \simeq 310$ MeV.

An interesting finite-temperature extension of the Weinberg sum rule for the
product of $\omega$ and the spectral density of vector mesons is discussed
in \cite{KAPU,ZSCH}. In the MF approximation resonances do not exist, 
so the constraints \cite{KAPU,ZSCH} apply starting from one loop ChPT. 

For $T < T_{max} = 3.36F \simeq 310$ MeV the system has regular behavior. The
scalar quark condensate decreases smoothly to zero. 
The effective pion mass increases with temperature in agreement with the
partial restoration of chiral symmetry and approaches ${\tilde M}_{\pi,max} = 1.61 {M}_{\pi}$. 
The pion optical potential $V_{opt}
= ({\tilde M}^2_{\pi} - M_{\pi}^2)/(2M_{\pi})$ increases also.

The pion optical potential $V_{opt}$ is calculated to all orders in
density. It corresponds to the summation of the forward scattering
amplitudes of a probing pion scattered on $n$ surrounding pions, with $n$
running from one to infinity.

For $T \simeq T_{max}$, $Z_{\pi}^{-1} \rightarrow \infty$. The pions disappear
from the spectral density of the propagator $<\mathcal{T}\pi^{\alpha}(x)\pi^{%
\beta}(y)> $. As elementary excitations, the pions do not disappear,
however, from energy spectrum Eq.(\ref{EDEN}) and contribute to the
pressure Eq.(\ref{PRES}). Moreover, the pions contribute to the spectral
density, e.g., of the two-point Green function $<\mathcal{T}\pi^{\alpha}(x)\chi^{%
\beta}(y)>$.

The estimate of $T_{max}$ is significantly above the chemical freeze-out temperature 
$T \simeq 170$ MeV at RHIC. 

It is worthwhile to notice that for massless pions, the scalar 
quark condensate vanishes at $T_{max}(M_{\pi} = 0) = 2.46F \simeq 230 $ MeV. The behavior of 
observables (\ref{M2EFF})-(\ref{FTILD}) does not change qualitatively. 
EOS becomes identical with EOS of the ideal pion gas, as can be seen from 
Eqs.(\ref{EDEN}) and (\ref{PRES}).

\begin{figure}[!htb]
\vspace{5mm}
\begin{center}
\includegraphics[width=0.381924\textwidth,height=0.236029032\textwidth]{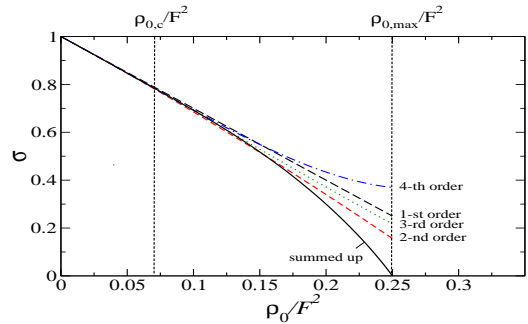}
\caption{The normalized scalar quark density 
$\sigma = <{\bar q}q>/<{\bar q}q>_{vac}$ versus the pion scalar density $\rho_{0}$.
The power series expansion (\ref{SCALA}) is truncated to orders $O(\rho_{0}^{n})$ 
for $n=1,2,3$ and $4$ (long-dashed, dashed, dotted, and dot-dashed curves, respectively) and results are compared to the exact numerical summation
of the series (\ref{SCALA}) using the contour-improved Borel resummation technique (solid curve).
The value of $\rho_{0,c}$ is the critical scalar pion density for phase transition into the 
quark matter. The value of $\rho_{0,max}$ is the maximum scalar pion density admissible 
in thermalized non-linear sigma model.
}
\label{consigma}
\end{center}
\end{figure}

Let us check accuracy of presentation of the results by truncated series. In Fig. \ref{consigma}, 
we show the ratio $\sigma = <{\bar q}q>/<{\bar q}q>_{vac}$ calculated using the series expansion (\ref{SCALA}) 
truncated to orders $O(\rho_{0}^{n})$ for $n=1,2,3$ and $4$. It is compared to the exact (numerical) summation of the perturbation series. We see that in the hadron phase $T \leq T_{c}$, the first order approximation is very precise. Noticeable, irregular deviations appear beyond $T_{c}$ only. We expect, therefore, a 20\% decrease of the scalar quark condensate at $T=T_{c}$.

The first-order approximation is sufficiently precise for ${\tilde M}_{\pi}^2$ also: At $T=T_{c}$, the first order 
of (\ref{M2EFF}) gives for the effective pion mass ${\tilde M}_{\pi}^2 \simeq 1.07 M_{\pi}^2$, while the exact summation gives $1.03 M_{\pi}^2$.
The other observables display the similar features. 


A prescription for approximate summation of asymptotic series, which is effective if initial sequential terms of asymptotic series decrease first before increasing, can be found in Ref. \cite{MIGD}. One should attempt to truncate asymptotic series where two sequential terms are of the same order. The magnitude of the first neglected term gives the accuracy of the summation. Comparing $n=1$ and $n=2$ terms of (\ref{SCALA}), one may conclude that accuracy of the series truncated at $n=1$ is 100\% for $\rho_{0} = 2F^2$. One can expect that for $\rho_{0,c} = 0.07F^2 \ll 2F^2$, the accuracy is very good. As we observed, this is the case. The same arguments require $\rho_{0}$ be less than $0.25F^2$ for ${\tilde M}_{\pi}^2$. Such a requirement is satisfied also, however, with a lower precision.

\section{Comparison with lattice gauge theories}
\setcounter{equation}{0} 

LGTs allow to calculate QCD observables from first principles. The smallness 
of physical quark masses and finite lattice spacing restrict the power of LGTs 
to temperatures above $\sim 100$ MeV. The non-linear sigma model, on the  other 
hand, is an effective theory of QCD at temperatures small 
compared to the pion mass. There are no intrinsic restrictions to extrapolate results of 
the non-linear sigma model up to $T_{max} \simeq 310$ MeV. However, above pion mass the 
non-linear sigma model represents in the strict sense a model rather than an effective theory of QCD. 

The phase transition to the quark matter appears in LGTs at $T_{c} \simeq 175$ MeV 
\cite{KALAPE,AALIK}. The domain of validity of the non-linear sigma model 
is therefore limited to temperatures below $T_{c}$. The interval of 
temperatures from $\sim 100$ MeV up to $T_{c}$ is 
suitable for a meaningful comparison of predictions from the non-linear sigma model 
and LTGs.

\begin{figure}[!htb]
\vspace{5mm}
\begin{center}
\includegraphics[width=0.381924\textwidth,height=0.236029032\textwidth]{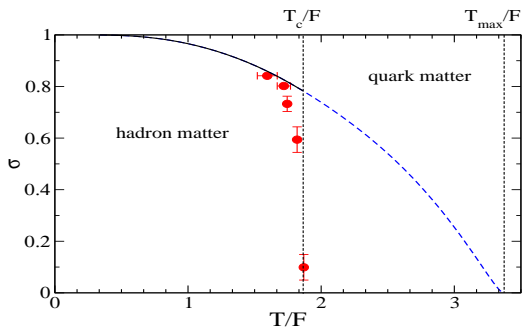}
\caption{The scalar quark condensate $\sigma = <{\bar q}q>/<{\bar q}q>_{vac}$ 
versus the temperature in units of $F=93$ MeV obtained within the non-linear sigma model
using the MF approximation (solid line below $T_{c}$ and dashed line above $T_{c}$) 
and from lattice gauge theory for two flavors 
\cite{KALA03}.
}
\label{scond}
\end{center}
\end{figure}

Fig. \ref{scond} shows the temperature dependence of the quark 
scalar condensate. The overall mass scale in LGTs is fixed assuming the string tension 
is flavor and quark mass independent. The agreement of the non-linear sigma model 
and LTG \cite{KALA03} is not unreasonable with regard to first 
two-three points with lowest temperatures. In the narrow deconfinement 
region the models strongly diverge, basically because 
the non-linear sigma model does not expose quark-gluon degrees of freedom.  

\begin{figure}[!htb]
\vspace{5mm}
\begin{center}
\includegraphics[width=0.381924\textwidth,height=0.236029032\textwidth]{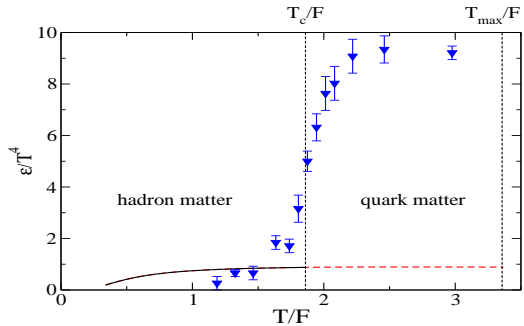}
\caption{The normalized energy density $\varepsilon/T^4$ versus the temperature $T$ 
obtained within the MF approximation of the non-linear sigma model (solid line below $T_{c}$ and 
dashed line above $T_{c}$) 
and from lattice gauge theory (filled triangles) with two flavors \cite{KALA03,KALAPE}.
The temperature is given in units of $F=93$ MeV.
}
\label{ed}
\end{center}
\end{figure}

The energy density as a function of temperature is shown in Fig. \ref{ed} at zero chemical potentials. The solid line and the dashed line are predictions of the non-linear sigma model. 
In addition, the lattice simulations from Refs. \cite{KALA03,KALAPE} are shown.

At a first-order phase transition, the energy density experiences a jump while
the pressure remains a smooth function. It is not quite clear from the LGT data 
presented on Fig. \ref{ed} whether we observe a crossover or a first-order phase transition. 
Usually, in the two-flavor case lattice simulations give 
evidence for a first-order phase transition, while pure gauge theory predicts 
a second-order transition and 2+1 flavor LGTs with physical quark masses predict 
a smooth crossover \cite{Aoki06}. 

If the phase transition is of first order , 
the energy density in LTG appears to be at $T_{c}$ several times 
greater than in the non-linear sigma model. Such a divergence can be attributed to 
resonances like $\rho$- and $\omega$-mesons which contribute to the energy density, 
but do not exist in the non-linear sigma model in MF approximation. An ideal 
resonance gas model which is in agreement with lattice simulations has e.g. been 
developed by Karsch et al. \cite{KART}. 

In the case of a first-order phase transition, the jump in the energy density can be 
as large as $\Delta\varepsilon_{c}/T^4_{c} \sim 8$, in which case pions alone 
saturate the energy density. If the jump is smaller, Hagedorn's conjecture on 
the exponential growth of the number of resonances with mass might be appropriate.
The value of the jump is crucial in order to understand the role of higher resonances.

Modelling the first-order phase transition \cite{KMK} within the framework of the MIT 
bag model \cite{WAIS} allows to determine the critical temperature using the pressure balance 
$p_{Q} = p_{H} + B$ where $B = 57$ MeV/Fm$^3$ is the vacuum pressure in 
$2+1$ flavor QCD \cite{MIT}, $p_{Q}$ and $p_{H}$ are pressures in quark and hadron phases, 
respectively, and determine jump in the energy density $\Delta \varepsilon_{c} = \varepsilon_{Q} + B - \varepsilon_{H}$ where $\varepsilon_{Q}$ and $\varepsilon_{H}$ are energy densities of quark matter and hadron matter, respectively. For ideal gases of relativistic particles, $\Delta \varepsilon_{c} = 3p_{Q} + B - 3p_{H} = 4B$, and so $\Delta \varepsilon_{c}/T^{4}_{c} \sim 3$.  
The vacuum pressure $B$ is density and temperature dependent \cite{SHUR,KOBZ,MULL,DILA,ASTR,STRA}, which brings to $\Delta \varepsilon_{c}$ more uncertainties. 

The energy density predicted by LTGs is obviously underestimated 
around $T \sim 100$ MeV due to unphysical quark masses $m_Q/T = 0.4$ and restrictions 
from finite lattice size. The energy density and the pressure depend strongly on the quark 
masses and the pion mass. The horizontal part of EOS at high $T$ (dashed line 
and partially solid line) is close to the energy density of 
massless pions $\varepsilon/T^4 = \pi^2/10$.

\begin{figure}[!htb]
\vspace{5mm}
\begin{center}
\includegraphics[width=0.381924\textwidth,height=0.236029032\textwidth]{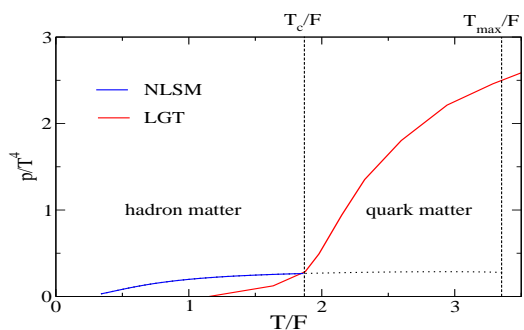}
\caption{The normalized pressure $p/T^4$ versus the temperature $T$ in units of $F=93$ MeV 
in the non-linear sigma model (NLSM) using the MF approximation 
(solid line below $T_{c}$ and dashed line above $T_{c}$)
and in lattice gauge theory (LGT)
with two flavors \cite{KALAPE,KALA03}.
}
\label{latt}
\end{center}
\end{figure}

The pressure is plotted on Fig. \ref{latt}. The lattice predictions \cite{KALAPE,KALA03} and the non-linear sigma 
model predictions are in agreement at the phase transition point, while at smaller temperatures 
the pressure in LGTs is significantly lower. The pressure of the pion gas 
approaches the pressure of ideal gas of massless 
pions $p/T^4 = \pi^2/30$ at $T \to T_{max}$. However, it  does not 
fully reach the ultrarelativistic limit, so the effects from the finite pion mass remain.

According to the Gibbs' criterion (see e.g. \cite{LANDAF}) the phase of matter 
with the highest presssure at equal temperatures and chemical potentials 
is preferable. At $T \sim T_{c}$, heavy resonances are still slowly moving 
particles which contribute to the energy density $\delta \varepsilon$ mainly through 
their rest mass $M$ 
and contribute to the pressure as $\delta p \sim \frac{T}{M} \delta \varepsilon  \ll \delta \varepsilon$. For zero chemical potentials, the 
pressure in hadron phase is dominated by light pions \cite{KMK}.

The non-linear sigma model predicts the critical pressure $p_{c}$ 
reasonably well. The critical
temperature can be obtained phenomenologically from 
Fig. \ref{latt} applying the Gibbs' criterion for the hadronic EOS derived using the non-linear 
sigma model and the quark matter EOS derived from  LGTs. The intersection of 
the two pressure curves yields then the temperature of the phase transition. 
The value obtained in such a way is in the remarkable agreement with 
lattice estimates based on the position of the steep rise of the energy 
density seen on Fig. \ref{ed}. 

\section{Conclusion}
\setcounter{equation}{0} 

It is generally believed that ChPT represents an adequate tool for studying
the pion matter at low temperatures. The Haar measure appearing in the path 
integral is important to keep the chiral symmetry at high temperatures 
unbroken within the MF approximation. In an arbitrary parameterization of the 
pion fields one faces a dilemma: Accounting for the effective potential 
$\delta \mathcal{L}_{H}$ arising due to the exponentiating the weigh factor of 
the path integral measure brings divergences which can be compensated by going beyond 
the MF approximation only, i.e., by including pion loops appearing  in the 
higher order ChPT expansion. If $\delta \mathcal{L}_{H}$ is neglected, the MF 
approximation does not restore the chiral invariance with increasing the 
temperature. There is only one parameterization with $\delta \mathcal{L}_{H} = 0$, 
which makes the MF approximation selfconsistent: The renormalizations are 
finite and the chiral symmetry is restored at high temperatures. This 
parameterization was used to study the in-medium modifications of pions 
and collective characteristics of the thermalized pion matter 
in the MF approximation. 

We made essentially two approximations:

(a) The results are obtained by extending the integration region over the pion fields
$\chi^{\alpha}$ from $- \infty$ to $+\infty$.

(b) The MF approximation consists in the neglection of loops 
composed from more than one dressed pion propagator. Loops formed 
by one dressed pion propagator give the pion density (\ref{PDEN}), i.e.,
the thermal parts of loops are accounted for, whereas the vacuum contributions 
of the zero-point field fluctuations are systematically neglected. 
An accounting for the vacuum parts of the pion loops in a thermal bath 
is possible beyond the lowest ChPT order.

We constructed Borel transforms of most important thermal averages. The corresponding
power series with respect to the density are asymptotic, have zero convergence 
radius, but are summable using the contour-improved Borel resummation method.. 
The pion gas does not exist above $T_{max} \simeq 310$ MeV, whereas 
at $T < T_{max}$ thermodynamic observables are smooth functions.

The deconfinement temperature $T_{c} = 1.86F$ and the corresponding scalar density 
$\rho_{0,c} = 0.07F^2$ appear to be small enough for approximate however quite accurate 
calculation of observables in hadron phase using truncated asymptotic series.

The region of validity of the method can be evaluated by requiring 
$<\chi^2> \leq \chi_{max}^2$. The corresponding restriction 
$T \leq \sqrt{8(3\pi)^{2/3}}F \approx 560$ MeV does not appear to be stringent.

The method of summation of the density series proposed in this work can 
be extended to higher orders of ChPT loop expansion.

\begin{acknowledgments}
M.I.K. wishes to acknowledge kind hospitality at the University of T\"ubingen. 
This work has been supported by DFG grant No. 436 RUS 113/721/0-2 and RFBR 
grant No. 06-02-04004.
\end{acknowledgments}

\begin{appendix} 

\section{MF approximation for $<({\chi }^{\gamma }{\chi }^{\gamma })^{n}>$}

The order of the operators $\chi ^{\alpha }=\chi ^{\alpha }(0,{\bf 0})$
entering the thermal average $<(\chi ^{\gamma }\chi ^{\gamma })^{n}>$ is not
specified so far. As we shall see, it is not important. Let us rewrite $%
<(\chi ^{\gamma }\chi ^{\gamma })^{n}>$ with the help of the ordering
operator ${\cal T}$ . To make its action well defined, we set first
arguments of the operators $\chi ^{\alpha }(0,{\bf 0})$ equal to small
parameters $\epsilon _{\alpha }$, i.e., we make replacements $\chi ^{\alpha
}(0,{\bf 0})\rightarrow \chi ^{\alpha }(\epsilon _{\alpha },{\bf 0})$. If
the limit $\epsilon _{\alpha }\rightarrow 0$ exists and does not depend on
the way the parameters $\epsilon _{\alpha }$ approach zero, the thermal
average $<(\chi ^{\gamma }\chi ^{\gamma })^{n}>$ can be expressed in terms
of the Green function: 
\begin{equation}
<(\chi ^{\gamma }\chi ^{\gamma })^{n}>=\lim_{\epsilon _{\alpha }\rightarrow
0}<{\cal T}\chi ^{\alpha _{1}}\chi ^{\alpha _{1}}\chi ^{\alpha _{2}}\chi
^{\alpha _{2}}...\chi ^{\alpha _{n}}\chi ^{\alpha _{n}}>.  \label{C2}
\end{equation}

In order to evaluate the Green function, we pass to the interaction
representation, 
\begin{equation}
<{\cal T}\chi ^{\alpha _{1}}\chi ^{\alpha _{1}}\chi ^{\alpha _{2}}\chi
^{\alpha _{2}}...\chi ^{\alpha _{n}}\chi ^{\alpha _{n}}>=<{\cal T}\chi
_{0}^{\alpha _{1}}\chi _{0}^{\alpha _{1}}\chi _{0}^{\alpha _{2}}\chi
_{0}^{\alpha _{2}}...\chi _{0}^{\alpha _{n}}\chi _{0}^{\alpha _{n}}S(\frac{1%
}{T},0)><S(\frac{1}{T},0)>^{-1}  \label{C3}
\end{equation}
where $\chi _{0}^{\alpha }(\tau ,{\bf x})=S(\tau ,0)\chi ^{\alpha }(0,{\bf x}%
)S^{-1}(\tau ,0)$ and $S(\tau _{2},\tau _{1})$ is the thermal $S$-matrix (see 
e.g. \cite{ABRI}). Applying the Wick's theorem to the $2n$-body
Green function (\ref{C3}) and neglecting all the diagrams except for those 
corresponding to the non-interacting gas of the dressed pions, we obtain 
\begin{equation}
{\cal G}(\epsilon _{1},\epsilon _{2},...,\epsilon _{2n})=\sum_{(\beta
_{1}\beta _{2}...\beta _{n}),(\gamma _{1}\gamma _{2}...\gamma \ _{n})}<{\cal %
T}\chi ^{\beta _{1}}\chi ^{\gamma _{1}}><{\cal T}\chi ^{\beta _{2}}\chi
^{\gamma _{2}}>...<{\cal T}\chi ^{\beta _{n}}\chi ^{\gamma _{n}}>  \label{C4}
\end{equation}
where $(\beta _{1}\beta _{2}...\beta _{n})$ and $(\gamma _{1}\gamma
_{2}...\gamma _{n})$ are permutations of $(\alpha _{1}\alpha
_{2}...\alpha _{n}).$ The summation runs over those permutations which occur
according to the diagram decomposition of the $2n$-body Green function of 
a non-interacting gas. The pion propagators are dressed. Entering Eq.(\ref{C4}) are
therefore the Green functions $<{\cal T}\chi ^{\alpha }\chi ^{\beta }>$
instead of $<{\cal T}\chi _{0}^{\alpha }\chi _{0}^{\beta }>$. 

\begin{figure}[!htb]
\begin{center}
\includegraphics[angle=0,width=3.2 cm]{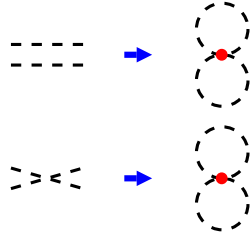}
\caption{
In the limit $\epsilon_{\alpha} \rightarrow 0$, the disconnected 
diagrams of the four-body Green function produce no loops with
more than one dressed pion propagator, being thus significant in the 
MF approximation.
}
\label{fig6}
\end{center}
\end{figure}

\begin{figure}[!htb]
\begin{center}
\includegraphics[angle=0,width=3.2 cm]{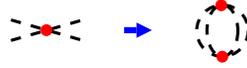}
\caption{
In the limit $\epsilon_{\alpha} \rightarrow 0$, connected diagrams of 
the four-body Green function produce loops involving more than one pion 
propagator and therefore are not significant in the MF approximation. 
The only diagrams essential for thermal average 
$<({\chi }^{\gamma }{\chi }^{\gamma })^{n}>$ are those which correspond 
to the non-interacting gas of the dressed pions.
}
\label{fig7}
\end{center}
\end{figure}

Show on Fig. \ref{fig6} are diagrams of the four-body Green function 
corresponding to the non-interacting gas of the effective pions. Such diagrams 
contribute to thermal averages. The diagram shown on Fig. \ref{fig7} 
which accounts for the pion rescattering should be discarded in the MF approximation.
This feature holds for $2n$-body Green functions.
The MF approximation does not involve terms except for those entering Eq.(\ref{C4}). 

The isotopic symmetry of the pion matter amounts to the two-body Green
function $\mathcal{G}^{\alpha \beta }(\tau ,{\bf k})$ which is symmetric under the
permutation $\alpha \leftrightarrow \beta $ at $\tau =0.$ Moreover, 
\begin{equation}
\lim_{\epsilon \rightarrow +0}\mathcal{G}^{\alpha \beta }(\epsilon ,{\bf k}%
)=\lim_{\epsilon \rightarrow -0}\mathcal{G}^{\alpha \beta }(\epsilon ,{\bf k}).
\label{UNIV}
\end{equation}
Eqs.(\ref{C4}) and (\ref{UNIV}) imply that after neglection by all terms proportional 
to $\rho_{vac}$ the limit $\epsilon _{\alpha
}\rightarrow 0$ in Eq.(\ref{C2}) exists indeed and, at least within the MF
approximation, does not depend on the way the parameters $\epsilon _{\alpha }
$ approach zero. Loops formed by one pion propagator give the
density (\ref{PDEN}), while loops involving 
more than one dressed pion propagator are neglected.

We obtain therefore 
\begin{equation}
<(\chi ^{\gamma }\chi ^{\gamma })^{n}>=(2\rho )^{n}\sum_{(\beta \ _{1}\beta
_{2}...\beta _{n}),(\gamma _{1}\gamma _{2}...\gamma _{n})}\delta ^{\beta
_{1}\gamma _{1}}\delta ^{\beta _{2}}{}^{\gamma _{2}}...\delta ^{\beta
_{n}\gamma _{n}}.  \label{C6}
\end{equation}
The combinatorial structure of the Wick's decomposition is identical with
that of the corresponding Gaussian integral, so one can write 
\begin{eqnarray}
&&\sum_{(\beta _{1}\beta _{2}...\beta _{n}),(\gamma _{1}\gamma _{2}...\gamma
\ _{n})}\delta ^{\beta _{1}\gamma _{1}}\delta ^{\beta _{2}}{}^{\gamma
_{2}}...\delta ^{\beta _{n}\gamma _{n}}  \nonumber \\
&=&\frac{1}{(2\pi )^{3/2}}\int \int \int x^{\alpha _{1}}x^{\alpha
_{1}}x^{\alpha _{2}}x^{\alpha _{2}}...x^{\alpha _{n}}x^{\alpha _{n}}\exp (-%
\frac{1}{2}((x^{1})^{2}+(x^{2})^{2}+(x^{3})^{2}))dx^{1}dx^{2}dx^{3} 
\nonumber \\
&=&\frac{4\pi }{(2\pi )^{3/2}}\int_{0}^{+\infty }r^{2n+2}\exp (-\frac{1}{2}%
r^{2})dr.  \label{C8}
\end{eqnarray}
Replacing $r\rightarrow \sqrt{2t},$ we arrive at Eq.(\ref{CALC}).

\section{Chiral invariance in the MF approximation}

As discussed in Sect. II, the MF approximation imposes restrictions on 
the pion field parameterizations. If the tree level $\delta \mathcal{L}_{H}$ 
is included into a MF calculation, one gets divergences. From other hand, 
the neglection by $\delta \mathcal{L}_{H}$ violates the chiral symmetry. 
Let us demonstrate it explicitly:

Equation (\ref{MAIN}) can be rewritten in more physical terms
\begin{equation}
<\mathcal{J}({\chi ^{2}})>=\frac{1}{(4\pi\rho)^{3/2}}\int {\mathcal{J}(\chi^2) e^{-\chi^2/(4\rho)}d^3\chi}.  \label{MAINP}
\end{equation}
The scalar product 
\begin{equation}
\cos (\Theta )=\frac{1}{2}{\mbox Tr}[U(\phi )U^{\dagger}(\phi ^{\prime })]
\end{equation} 
is obviously chirally invariant. The value of $\Theta =\Theta (\phi,\phi ^{\prime })$ determines 
the angular distance between two sets of the pion fields. The parameter space is thus a metric 
space with an infinitesimal distance 
\begin{equation}
d\Theta ^{2}=d\phi ^{2}+\sin ^{2}(\phi )(d\theta ^{2}+\sin ^{2}(\theta)d\varphi ^{2})
\end{equation}
where $\theta $ and $\varphi$ are polar 
and azimuthal angles of the vector ${\phi }^{\alpha }$. 

Equation (\ref{MAINP}) has a transparent physical meaning. Quantum fluctuations
of the fields $\chi^{\alpha}$ induce fluctuations of a function $\mathcal{J}({\chi ^{2}})$.
These fluctuations are suppressed by the exponential factor entering (\ref{MAINP}). 

According to Eq.(\ref{CHI}) $\chi$ depends on $\phi$ which has the meaning of a metric distance
\begin{equation}
\phi = \Theta(\phi^{\alpha},\phi^{\alpha}_{vac})
\end{equation}
between the vector $\phi^{\alpha}$ and the vector $\phi^{\alpha}_{vac}=0$ which specifies 
the vacuum state on the $4$-dimensional sphere, making thereby the chiral symmetry spontaneously broken. The only essential place in Eq.(\ref{MAINP}) where the dependence on the $\phi^{\alpha}_{vac}=0$ shows up is the 
exponential factor. In the limit $T \rightarrow \infty$ the density increases, $\rho \rightarrow \infty$, 
and so the dependence on $\phi^{\alpha}_{vac}=0$ drops out. The exponential factor becomes a constant.
It means that quantum fluctuations at all points of the $4$-dimensional sphere contribute equally to 
thermal averages. This is in agreement with our expectations that the chiral 
symmetry restores with increasing the temperature. 

A use of the exponential parameterization with the Haar measure neglected would result in
\begin{equation}
<\mathcal{J}({\chi ^{2}})> {\stackrel{?}{=}} \frac{1}{(4\pi\rho)^{3/2}}\int {\mathcal{J}(\chi^2) e^{-\phi^2/(4\rho)}d^3\phi}.  \label{MAINW}
\end{equation}
The Euclidean volume $d^3 \phi $ keeps explicitly a reference to the vacuum vector $\phi_{vac}^{\alpha}=0$. 
It means that the high temperature limit does not restore the chiral invariance in the MF approximation if the Haar measure is neglected.

\end{appendix}


\begin{thebibliography}{99}

\bibitem{KALAPE} F. Karsch, E. Laermann and A. Peikert, Nucl. Phys. {\bf B605}, 579 (2001).

\bibitem{AALIK} A. Ali Khan {\it et al.}, Phys. Rev. {\bf D63}, 034502 (2001).


\bibitem{Aoki06}
  Y.~Aoki, Z.~Fodor, S.~D.~Katz and K.~K.~Szabo,
  arXiv:hep-lat/0609068.

\bibitem{BRMU01}  P. Braun-Munzinger, D. Magestro, K. Reidlich, and J.
Stachel, Phys. Lett. \textbf{B518}, 41 (2001).

\bibitem{andronic06} A. Andronic, P. Braun-Munzinger, J. Stachel, Nucl. Phys.. {\bf A772}, 167 (2006).

\bibitem{BRAV02}  L. Bravina, A. Faessler, C. Fuchs, E. E. Zabrodin,
Z.-D. Lu, Phys. Rev. \textbf{C66}, 014906 (2002).

\bibitem{KMK} L. A. Kondratyuk, B. V. Martemyanov, M. I. Krivoruchenko, Z. Phys. {\bf C52}, 563 (1991).

\bibitem{KART} F. Karsch, K. Redlich, A. Tawfik, Eur. Phys. J. {\bf C29}, 549 (2003).

\bibitem{hagedorn1} R. Hagedorn, Nuovo Cim. {\bf 35}, 395 (1965).

\bibitem{hagedorn2} R. Hagedorn, in Cargèse Lectures in Physics, Vol. 6, edited by E. Schatzman (Gordon and Breach, New York, 1973).

\bibitem{KALA03} F. Karsch and E. Laermann, e-Print Archive: hep-lat/0305025..


\bibitem{GL84}  J.~Gasser and H.~Leutwyler, Ann. Phys.\ \textbf{158}, 142
(1984).

\bibitem{GL85}  J.~Gasser and H.~Leutwyler, Nucl. Phys. \textbf{B250}, 465
(1985).

\bibitem{MEIS}  U. G. Meissner, Rep. Progr. Phys. \textbf{56}, 903 (1993).

\bibitem{DOBA}  A. Dobado, A. N. G\'{o}mez, A. L. Maroto, and J. R.
Pel\'{a}ez, \textit{Effective Lagrangians for the Standard Model},
Springer-Verlag, Berlin (1997).


\bibitem{GL87}  J. Gasser and H. Leutwyler, Phys. Lett. \textbf{B184}, 83
(1987).

\bibitem{GELE89}  P. Gerber and H. Leutwyler, Nucl. Phys. \textbf{B321}, 387
(1989).

\bibitem{ELET93}  V. L. Eletsky, Phys. Lett. \textbf{B299}, 111 (1993).

\bibitem{ELET94}  V. L. Eletsky and Ian I. Kogan, Phys. Rev. \textbf{D49},
3083 (1994).

\bibitem{PITY96}  R. D. Pisarski and M. Tytgat, Phys. Rev. \textbf{D54},
R2989 (1996).

\bibitem{BOKA96}  A. Bochkarev and J. Kapusta, Phys. Rev. \textbf{D54},
4066 (1996).

\bibitem{JEON}  S. Jeon, J. Kapusta, Phys. Rev. \textbf{D54}, R6475 (1996).

\bibitem{TOUB97}  D. Toublan, Phys. Rev. \textbf{D56}, 5629 (1997).

\bibitem{dobado}  A. Dobado, J. R. Pel\'{a}ez, Phys. Rev. \textbf{D59},
034004 (1999).

\bibitem{KRIV}  B. V. Martemyanov, A. Faessler, C. Fuchs, M. I.
Krivoruchenko, Phys. Rev. Lett. \textbf{93}, 052301 (2004).

\bibitem{LOEWE}  M. Loewe, C. Villavicencio, Phys. Rev. \textbf{D71}, 094001 (2005).


\bibitem{CHIK}  J. Chisholm, Nucl. Phys. \textbf{26}, 469 (1961).

\bibitem{KAME}  S.~Kamefuchi, L.~O'Raifeartaigh, and A.~Salam, 
Nucl.\ Phys.\ \textbf{28}, 529 (1961). 

\bibitem{Tho95}  V. Thorsson and A. Wirzba, Nucl. Phys. \textbf{A589}, 633
(1995).

\bibitem{Lee95}  C.-H. Lee, G. E. Brown, D.-P. Min, and M. Rho, Nucl. Phys. 
\textbf{A585}, 401 (1995).

\bibitem{Par02}  T.-S. Park, H. Jung, D.-P. Min, J. Kor. Phys. Soc. \textbf{%
41}, 195 (2002).

\bibitem{Kon03}  S. Kondratyuk, K. Kubodera, F. Myhrer, Phys. Rev. \textbf{%
C68}, 044001 (2003).

\bibitem{WIGN}  E. Wigner, \textit{Group Theory}, Academic, New York, 1959, p. 152.

\bibitem{MIK}  M. I. Krivoruchenko, A. Faessler, A. A. Raduta, C. Fuchs,
Phys. Lett. \textbf{B608}, 164 (2005).

\bibitem{WEISS} N. Weiss, Phys. Rev. {\bf D24}, 475 (1981). 

\bibitem{SAIL}  K. Sailer, A. Schafer, W. Greiner, Phys. Lett. {\bf B350}, 234 (1995). 

\bibitem{PWEIN} S. Weinberg, Phys. Rev. {\bf 166}, 1568 (1968).


\bibitem{DOWK} J. S. Dowker, J. Phys. {\bf A} 3, 451 (1970); Ann. Phys. (N.Y..) {\bf 62}, 361 (1971).
\bibitem{MATE} M. S. Marinov and M. V. Terent'ev, Yad. Fiz. {\bf 28}, 1418 (1978) [Sov. J. Nucl. Phys. {\bf 28}, 729 (1978)]; \\ Fortschr. Phys. {\bf 27}, 511 (1979).
\bibitem{SCHU} L. Schulman, Phys. Rev. {\bf 176}, 1558 (1968).
\bibitem{DURU} I. H. Duru, Phys. Rev. {\bf D30}, 2121 (1984).
\bibitem{BAAQ} B. E. Baaquie, Phys. Rev. {\bf D32}, R1007 (1985).


\bibitem{ABRI}  A. A. Abrikosov, L. P. Gorkov, I. E. Dzyaloshinsky, \textit{%
Methods of Quantum Field Theory in Statistical Physics}, Prentice-Hall, Inc.
Englewood Cliffs, N. J. (1963).

\bibitem{DYSON}  F. J. Dyson, Phys. Rev. \textbf{85}, 613 (1952).

\bibitem{ITZU}  C. Itzykson and J.-B. Zuber, \textit{Quantum Field Theory},
McGraw-Hill, Inc. New York, 1980, Chap. 9-4.

\bibitem{BEND}  C. M. Bender and S. A. Orszag, \textit{Advanced Mathematical
Methods for Scientists and Engineers}, McGraw-Hill, New York (1978).

\bibitem{JEAN}  U. D. Jentschura, Phys. Rev. \textbf{A64}, 013403 (2001).

\bibitem{KAPU}  J. I. Kapusta and E. V. Shuryak, Phys. Rev. \textbf{D49},
4694 (1994).

\bibitem{ZSCH}  S. Zschocke, O. P. Pavlenko, B. Kampfer, Eur. Phys. J. 
\textbf{A15}, 529 (2002).

\bibitem{MIGD} A. B. Migdal, {\it Qualitative Methods in Quantum Theory}, W. A. Benjamin, Inc., Massachusetts (1977).




\bibitem{WAIS} A. Chodos, R. L. Jaffe, K. Johnson, C. B. Thorn, V. F. Weisskopf, Phys. Rev. {\bf D9}, 3471 (1974).

\bibitem{MIT} T. A. DeGrand, R. L. Jaffe, K. Johnson, J. E. Kiskis, Phys. Rev. {\bf D12}, 2060 (1975).

\bibitem{SHUR} E. V. Shuryak, Phys. Lett. {\bf 79}, 135  (1978).

\bibitem{KOBZ} I. Yu. Kobzarev, B. V. Martemyanov, M. G. Shchepkin, Yad. Fiz. {\bf 29}, 1620 (1979)
[Sov. J. Nucl. Phys. {\bf 29}, 831 (1979)].

\bibitem{MULL} B. Muller, J. Rafelski, Phys. Lett. {\bf B101}, 111 (1981).

\bibitem{DILA} L. A. Kondratyuk, M. I. Krivoruchenko, M. G. Shchepkin, 
Pis'ma v ZHETF {\bf 43}, 10 (1986) [JETP Lett. {\bf 43}, 10 (1986)];
Yad. Fiz. {\bf 45}, 514 (1987) [Sov. J. Nucl. Phys. {\bf 45}, 323 (1987)]. 

	
\bibitem{ASTR} L. A. Kondratyuk, M. I. Krivoruchenko, B. V. Martemyanov, 
Pis'ma v Astron. Zh. {\bf 16}, 954 (1990) [Sov. Astron. Lett. {\bf 16}, 410 (1990)].

\bibitem{STRA} N. Prasad, R. S. Bhalerao, Phys. Rev. {\bf D69}, 103001 (2004).

\bibitem{LANDAF}  L. Landau and E. Lifshitz, {\it Statistical Mechanics}, Pergamon, New York, (1958).


\end{thebibliography}
\end{document}